\documentclass[twocolumn,showpacs,pra,amsmath,amssymb]{revtex4}
\usepackage{amsfonts, amssymb,amsmath}
\usepackage{graphicx}
\usepackage{dcolumn}
\usepackage{bm}

\begin{document}

\title{Superluminal and Ultraslow Light Propagation in Optomechanical Systems}
\author{Devrim Tarhan$^{1}$, Sumei Huang$^{2}$, and \"Ozg\"{u}r E. M\"ustecapl{\i}o\u{g}lu$^3$}
\affiliation{$^1$ Department of Physics, Harran University, Osmanbey Yerle\c{s}kesi,
\c{S}anl\i{}urfa, 63300, Turkey}

\affiliation{$^2$ Department of Physics, University of California,
Merced, California 95343, USA}

\affiliation{$^3$Department of Physics, Ko\c{c} University,
Sar{\i}yer, Istanbul, 34450, Turkey}

\email{dtarhan@harran.edu.tr}
\date{\today}
\begin{abstract}
We consider an optomechanical double-ended cavity under the action of
a coupling laser and a probe laser in electromagnetically induced transparency configuration.
It is shown how the group delay and advance of the probe field can be controlled
by the power of the coupling field. In contrast to single-ended cavities, only allowing for superluminal
propagation, possibility of both superluminal
and subluminal propagation regimes are found. The magnitudes of the group delay and the
advance are calculated to be $\sim 1\,$ms and $\sim -2\,$s, respectively, at a
very low pumping power of a few microwatts. In addition, interaction of the
optomechanical cavity with a time dependent probe field is investigated
for controlled excitations of mirror vibrations.
\end{abstract}
\pacs{42.50.Gy, 42.50.Ct, 42.50.Wk} \maketitle
%
%
%
\section{Introduction} \label{sec:introduction}
The demonstration of ultraslow group velocity ($v_g$) of light
\cite{slowlight-exp1} in ultracold atoms by electromagnetically induced transparency
(EIT) \cite{eit1} has inspired appealing applications
\cite{slowlight-apps1,slowlight-apps2,slowlight-apps3,slowlight-apps4,qmemory1,lightstorage}.
Besides the slow light, superluminal phenomena ($ v_g > c \,\, or
\, \, v_g \, \,$ is negative) was observed in atomic caesium gas
\cite{kuzmich,kuzmich1} and in alexandrite crystal
\cite{bigelow}. Slow and superluminal light have also been observed in
optomechanical systems \cite{painter} whose
superior delay and advancement times, smaller dimensions,
and less demanding thermal requirements makes them attractive for quantum optomechanical
memory and classical signal processing applications
\cite{painter1,favero,schliesser,aspel,thompson,meystre,kippenberg,nunnenkamp,agarwal1,agarwal2012}.
Recent proposal such as
 optomechanical cavity with a Bose-Einstein condensate (BEC) \cite{kadizhu} and one-sided
cavity with a nanomechanical mirror (NMM) \cite{agarwal2}, which is recently demonstrated \cite{kippenberg},
are promising but either too costly and difficult to implement \cite{kadizhu}  or not sufficiently flexible enough
to realize both superluminal and slow light effects simultaneously  \cite{agarwal2,kippenberg}. The analogue of electromagnetically induced transparency has been demonstrated very recently in a room temperature cavity optomechanics setup formed by a thin semitransparent membrane within a Fabry-Perot cavity \cite{vitali}.
We address the question of
how more controllable and simpler optomechanical
systems, that can simultaneously exhibit larger delay and advancement
times, can be realized.

In this work,  we investigate the time delay of the weak probe
field at the probe resonance in a double-ended high quality cavity with a moving
NMM under the action of coupling laser. We find that the group delay can be controlled by
the power of the coupling field. The time delay is positive which
corresponds to ultraslow light propagation (subluminal
propagation) when there is a strong coupling between the
nano-oscillator and the cavity. In contrast to single-ended cavities, only allowing for superluminal
propagation \cite{tarhan}, possibility of both superluminal
and subluminal propagation regimes are found. The magnitude of the group delay
is $\sim 2\,$ms at a very low pumping power of a few
microwatts. The transmission group delay that we have found is
larger than the group delay in a coupled BEC-cavity system
\cite{kadizhu} which is costly and difficult system to implement.
In addition,
we show that it is possible to control the vibrational excitations
of the NMM by time dependent probe field.

Organization of the paper is as follows. In Sec. \ref{sec:model} we describe
our physical optomechanical system and EIT configuration. The quantities
such as group delay and advancement times redefined here as well. Results are
given in Sec. \ref{sec:results} in two sub-sections. The first sub-section is
dedicated to the case of constant pump and probe fields while the other one
focuses on the case of time-dependent fields. Conclusion is given in Sec. \ref{sec:conclusion}.
\section{Model System} \label{sec:model}
\begin{figure}[!t]
\begin{center}
\includegraphics[width=0.4\textwidth]{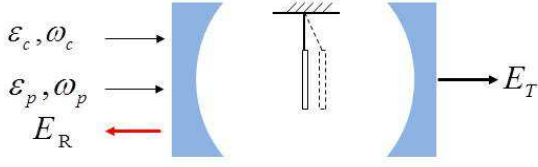}
\caption{\label{fig1}(Color online) Schematic of a double-ended cavity with a moving
nanomechanical mirror adapted from Ref. {\cite{agarwal1}}.}
\end{center}
\end{figure}

We consider the classical probe
field $\varepsilon_p$  and calculate the
response of the cavity optomechanical system to the probe field
 in the presence of the coupling field
$\varepsilon_c$. The
nanomechanical oscillator of frequency $\omega_m$ is coupled to a
Fabry-Perot cavity via radiation pressure effects \cite{agarwal1}.
In a Fabry-Perot cavity, both mirrors have equal reflectivity. We
use a configuration in which a partially transparent NMM is in the middle of a cavity
that is bounded by two high-quality mirrors as shown in Fig. \ref{fig1}.
The system is driven by a coupling
field of frequency $\omega_c$ and the probe field has frequency
$\omega_p$.
The Hamiltonian of this system is given by
\begin{eqnarray} \label{Ham}
H  &=& \hbar (\omega_0 - \omega_c)  c^{\dagger} c + \hbar g
c^{\dagger} c q + \frac{p^2}{2m}+\frac{1}{2}m \omega_m^2 q^2
\nonumber \\ &+& i\hbar \varepsilon_c (c^{\dagger} - c ) + i\hbar
(c^{\dagger}\varepsilon_p e^{-i\delta t}-c \varepsilon_p^\ast
e^{i\delta t}),
\end{eqnarray}
where $\delta=\omega_p-\omega_c$, $g=-\omega_c/L$ is the coupling constant between the
cavity field and the movable mirror \cite{meystre1},  and $c,c^{\dagger}$ are the
annihilation creation and  operators of the photons of the cavity field respectively.
The momentum and position operators of the
nanomechanical oscillator are $p$ and $q$, respectively. The amplitude of the pump
field is $\varepsilon_c=\sqrt{2\kappa P_c/\hbar \omega_c}$ with
$P_c$ being the pump power.

Heisenberg equation of motion for the coupled cavity-mirror system
is written and the damping rate  $2\kappa$ is added
phenomenologically  to represent the loss at the cavity
mirrors.The system is examined in the mean field limit
\cite{agarwal2}
\begin{eqnarray}\label{new}
\langle \dot{q}\rangle&=&\frac{\langle p\rangle}{m},\nonumber\\
\langle \dot{p}\rangle&=&-m\omega_{m}^2\langle q\rangle-\hbar g\langle c^{\dag}\rangle\langle c\rangle-\gamma_{m}\langle p\rangle,\nonumber\\
\langle \dot{c}\rangle&=&-[2\kappa+i(\omega_{0}-\omega_{c}+g\langle q\rangle)]\langle c\rangle+\varepsilon_{c}+\varepsilon_{p}e^{-i\delta t},\nonumber\\
\langle \dot{c}^{\dag}\rangle&=&-[2\kappa-i(\omega_{0}-\omega_{c}+g\langle q\rangle)]\langle c^{\dag}\rangle+\varepsilon_{c}+\varepsilon_{p}^{*}e^{i\delta t}.
\end{eqnarray}
The linear response solution is developed analytically using the ansatz \cite{boyd},
\begin{eqnarray}
q(t)  &=& q_0 + q_+ \varepsilon_p e^{-i \delta t} + q_- \varepsilon_p^* e^{i \delta t}, \nonumber \\
p(t)  &=& p_0 + p_+ \varepsilon_p e^{-i \delta t} + p_- \varepsilon_p^* e^{i \delta t},  \label{anzat} \\
c(t)  &=& c_0 + c_+ \varepsilon_p e^{-i \delta t} + c_- \varepsilon_p^* e^{i \delta t}, \nonumber
\end{eqnarray}
where $q_0,p_0$ and $c_0$ are the zeroth order solutions while the next terms corresponds to the
first order solutions in probe field amplitude.
By inserting Eq. \ref{anzat} into the Heisenberg equation of motion
we first obtain the steady state solutions
$c_0 = \varepsilon_c/(2 \kappa+i \Delta)$ and $ q_0
= - \hbar g |c_0|^2/m \omega_m^2$. Using them the first order solutions are
analytically determined to be
\begin{eqnarray} \label{cplus}
c_+(\delta) =  \frac{ m(\delta^2-\omega_m^2 + i \gamma_m \delta)[2
\kappa - i(\Delta + \delta)] - i \alpha }{m(\delta^2-\omega_m^2 +
i \gamma_m \delta)[(2\kappa - i \delta)^2 + \Delta^2] + 2 \Delta
\alpha },
\end{eqnarray}
where $\Delta=\omega_0 - \omega_c + gq_0 $ is the effective
detuning and $\alpha=\hbar g^2 |c_0|^2$. $|c_0|^2$ is the resonator intensity
and $q_0$ is the steady state position of
the movable mirror.

We can write the output field
$\varepsilon_{out}(t)  = \varepsilon_{out0} + \varepsilon_{out+}
\varepsilon_p e^{-i \delta t} + \varepsilon_{out-} \varepsilon_p^*
e^{i \delta t}$ \cite{boyd}. Inserting this to the input-output
relation and  comparing the coefficient of $\varepsilon_{p}e^{-i\delta t}$, we get the probe response $(
\varepsilon_{out+} + 1)=2 \kappa c_+$. The reflection and the
transmission of the probe response are denoted by $\varepsilon_R$
and $\varepsilon_T$, respectively. The reflection and transmission of the output
field respectively are determined by $E_{R}=\varepsilon_{R}\varepsilon_p e^{-i \omega_p t}$
and
$E_{T}=\varepsilon_{T}\varepsilon_p e^{-i \omega_p t}$. $\varepsilon_{T}=2\kappa c_+(\delta) $ is the transmitted
and $\varepsilon_{R}=2\kappa c_+(\delta) - 1 $ is the reflected components
of the probe field.  The amplitude
of the transmission output field is
$E_{T} = |T| \varepsilon_p \exp{({\textrm i}\phi( \omega_p))}$.

If we expand $\phi(\omega_p)$ around $\overline{\omega}$ to the first order
\begin{equation} \label{phase1}
\phi(\omega_p) = \phi(\overline{\omega}) + (\omega_p -
\overline{\omega}) \frac{\partial \phi}{\partial \omega_p} \mid
_{\overline{\omega}},
\end{equation}
the transmitted probe pulse can be expressed as $|T|
\,\varepsilon_p e^{-i \omega_p t} e^{i
\phi(\overline{\omega})}e^{i (\omega_p -
\overline{\omega})\frac{\partial \phi}{\partial \omega_p} \mid
_{\overline{\omega}}}$, where $\phi(\overline{\omega})=0$ at
resonance. Combining with the $e^{-i \omega_p(t - \tau)}$, the
transmitted probe pulse peaks at $t=\tau$, where $\tau$ is the
pulse delay that can defined as
\begin{equation}
\tau =  [ \frac{\partial \phi}{\partial \omega_p} ] \mid
_{\overline{\omega}}.
\end{equation}
Phase of the output field can be found as
\begin{equation} \label{phase2}
\phi = \frac{1}{2i} \ln(
\frac{\varepsilon_{T}}{\varepsilon_{T}^\star}).
\end{equation}
The time delay of the transmission and reflection pulse can be
determined by
\begin{eqnarray}
\tau_T = \texttt{Im} [\frac{1}{\varepsilon_{T}}\frac{\partial
 \varepsilon_{T}}{\partial \omega_p}] |_{\overline{\omega}}
 \nonumber \\
 \tau_R = \texttt{Im} [\frac{1}{\varepsilon_{R}}\frac{\partial
 \varepsilon_{R}}{\partial \omega_p}] |_{\overline{\omega}}
\end{eqnarray}
%
\section{Results and discussion} \label{sec:results}
In our calculations, we use parameters \cite{thompson} for the
length of the cavity $L=6.7$ cm, the wavelength of the laser
$\lambda=2\pi c/\omega_c=1064$ nm, $m=40$ ng, $\omega_m=2\pi
\times 134$ kHz, $\gamma=0.76$ Hz, $\kappa=\omega_m/10$ and
mechanical quality factor $Q=1.1 \times\times 10^6$, and
$\Delta=\omega_m$. The real and the imaginary parts of the
($\varepsilon_{T}=2 \kappa c_+$) represent the absorptive and
dispersive behavior, respectively.
\subsection{Constant Pump and Probe Fields} \label{subsec:constantFields}
\begin{figure}[!t]
\centering{\vspace{0.5cm}}
\includegraphics[width=0.4\textwidth]{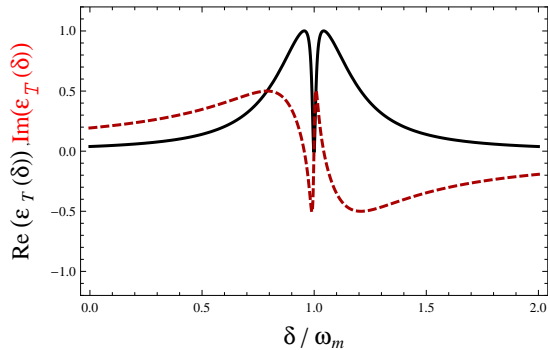}
\caption{\label{fig2}(Color online) The the real (black solid) and the imaginary (red dashed) parts of the
$\varepsilon_{T}(\delta)$ as a function of $\delta$ for input
coupling laser power of $P_c = 5 \mu$ W. The parameters used are
the length of the cavity $L=6.7$ cm, the wavelength of the laser
$\lambda=2\pi c/\omega_c=1064$ nm, $m=40$ ng, $\omega_m=2\pi
\times 134$ kHz, $\gamma=0.76$ Hz, $\kappa=\omega_m/10$ and
mechanical quality factor $Q=1.1 \times 10^6$, and
$\Delta=\omega_m$. }
\end{figure}
We show the real and the imaginary parts of the $\varepsilon_{T}$ In Fig. \ref{fig2}.
Under the conditions of electromagnetically
induced transparency in the mechanical system contained in a high
quality cavity the system gives rise to dispersion that leads to ultraslow propagation of the probe
field \cite{agarwal1}.
\begin{figure}[!t]
\begin{center}
\includegraphics[width=0.4\textwidth]{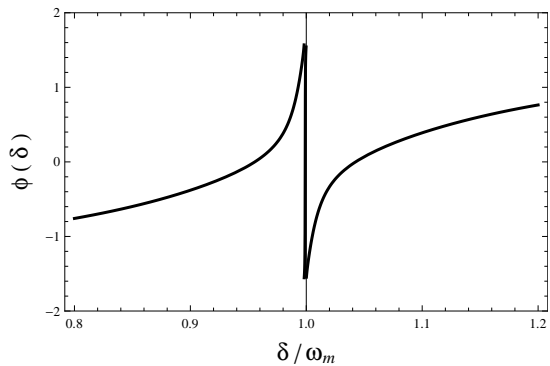}
\caption{\label{fig3} Phase as a function of frequency $\delta$ for input
coupling laser power $P_c = 1\mu$ W. The parameters are the same
as Fig. \ref{fig2}.}
\end{center}
\end{figure}
The phase is determined by Eq. \ref{phase2} and $\varepsilon_{T}$, and the result is plotted
in Fig. \ref{fig3} as a function of the scaled dimensionless
frequency $\delta / \omega_m$ for the input coupling laser power $P_c
= 1 \mu$W.

\begin{figure}[!t]
\begin{center}
\includegraphics[width=0.4\textwidth]{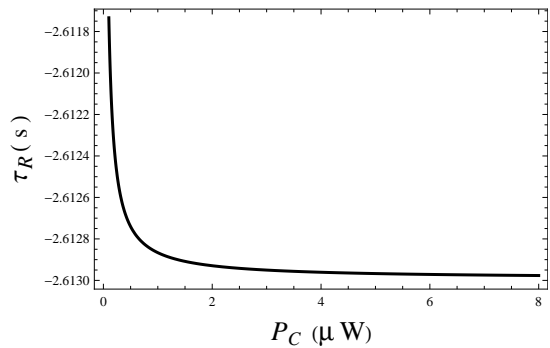}
\caption{ \label{fig4} Advance of the reflected probe field as a function of the pump power in the presence of the coupling field. All parameters are the same as those in Fig.
\ref{fig2}.}
\end{center}
\end{figure}

\begin{figure}[!t]
\begin{center}
\includegraphics[width=0.4\textwidth]{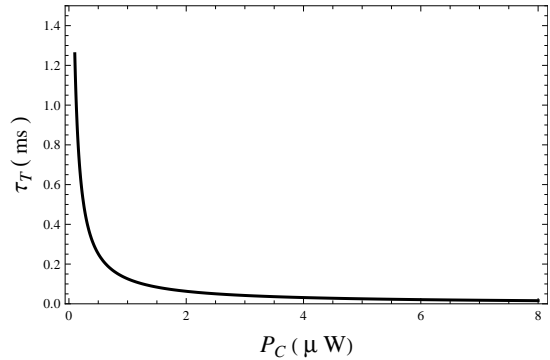}
\caption{\label{fig5} Group delay of the transmitted probe field as a function
of the pump power in the presence of the coupling field. All parameters are the same as those in
 Fig. \ref{fig2}.}.
\end{center}
\end{figure}

In the case of  no coupling field  $g=0$, the delay
time becomes $\tau_0=1.48 \, \mu$s. The coupling
reverses the behavior of the system and the group delay becomes
positive. We plot the group delay $\tau$ as a function of the pump
power in Fig. \ref{fig4} and Fig. \ref{fig5} which show the group
delay $\tau$ as a function of the pump power $P_c$. The group
delay decreases with increasing power of the coupling field. The
probe pulse delay can be tuned by calibrating the pump power in
the probe resonance ($\delta= \omega_m$).
The pump power that we have used in Fig. \ref{fig4} and Fig.
\ref{fig5} is on the order of $0.1-5 \mu$W. In Fig.
\ref{fig4}, the group delays are negative, which means that the reflected probe field is a superluminal light.
%
%
%
%
In Fig. \ref{fig5} the group
delays are positive, as a result the slow light effect can be
observed. This corresponds to a subluminal situation. We find large positive group delays of order $2$ ms in a
Fabry-Perot cavity under the action of a coupling laser and a
probe laser. The physics of subluminal or superluminal light propagation in double-ended cavity optomechanical system is associated with the interaction of NNM and cavity field.

\begin{figure}[!t]
\begin{center}
\includegraphics[width=0.4\textwidth]{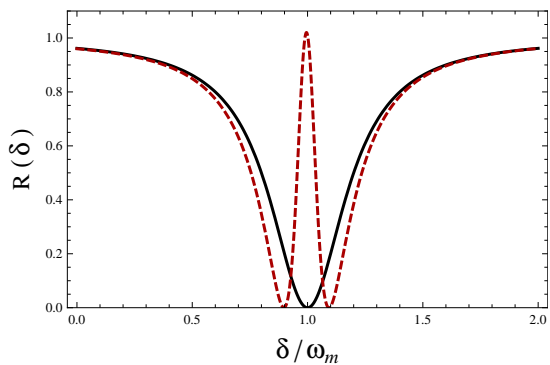}
\caption{\label{fig6} (Color online) The reflection spectrum $R(\delta)$ as a function of
normalized frequency. $P_c=0$ solid,
5$\mu$W (dashed). All parameters are the same with those of Fig.
\ref{fig2}.}.
\end{center}
\end{figure}

\begin{figure}[!t]
\begin{center}
\includegraphics[width=0.4\textwidth]{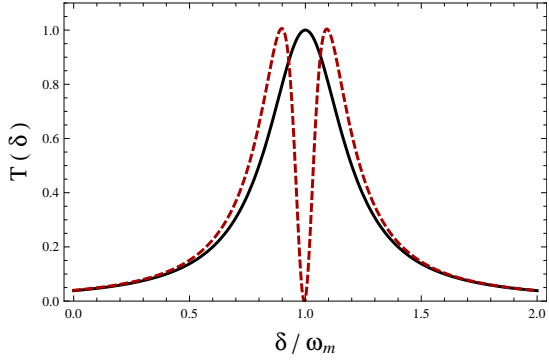}
\caption{\label{fig7} (Color online) The transmission spectrum
$T(\delta)$ as a function of normalized frequency. $P_c=0$ solid,
5$\mu$W (dashed). All parameters are the same with those of Fig.
\ref{fig2}.}.
\end{center}
\end{figure}

\begin{figure}[!t]
\begin{center}
\includegraphics[width=0.4\textwidth]{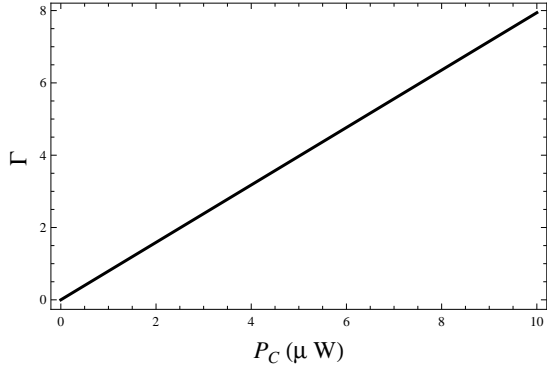}
\caption{\label{fig8} EIT width  $\Gamma(P_c)$ as a
function of power $P_c$. $\Gamma$ is normalized by $1\times 10^4$ 1/s.  All parameters are the same with those of
Fig. \ref{fig2}.}.
\end{center}
\end{figure}

We plot the reflection $R(\delta)=|2\kappa c_+ - 1|^2$ and
transmission spectrums $T(\delta)=|2\kappa c_+ |^2$ of the probe
field
 respectively.  in Fig. \ref{fig6} and Fig. \ref{fig7}.
%
%
The width of the transparency window of EIT is given by \cite{agarwal1}:
\begin{equation} \label{width}
\Gamma(P_c) = \frac{\gamma_m}{2} + \frac{\alpha(P_c)}{4m \omega_m
\kappa},
\end{equation}
where $\alpha(P_c)=\hbar g^2 \mid c_0\mid^2$. EIT width changes with the power in linear manner as shown in
Fig. \ref{fig8}.
%
\subsection{Time Dependent Probe Field} \label{subsec:timeDependetFields}
We now consider the time dependent, pulsed, probe field. As the
dynamics of the optomechanical system is associated with the
normal modes of the cavity field and the mirror vibrations, it is
natural to expect the oscillation modes of the mirror can be
controlled with temporal profile of the optical fields. We examine
particularly pulses with duration much less than the
characteristic time of mirror oscillations. The effect of such
pulses can be interpreted as if the mirror oscillator is kicked by
the optical pulses in sudden perturbations. We find the situations
of both robust excitations of mirror motional modes where the
mirror is simply displaced without oscillations and the periodical
excitations where the mirror vibrates.

Now we use the following ansatz in terms of time dependence in
order to obtain resonator-mirror coupled equations:
\begin{eqnarray}\label{anzat1}
q(t)  &=& q_0(t) + q_+(t) e^{-i \delta t} + q_-(t) e^{i \delta t},  \nonumber \\
p(t)  &=& p_0(t) + p_+(t) e^{-i \delta t} + p_-(t) e^{i \delta t},  \\
c(t)  &=& c_0(t) + c_+(t) e^{-i \delta t} + c_-(t) e^{i \delta t}.
\nonumber
\end{eqnarray}
Since the probe field is very weak, the force acting on the mirror exerted by the time-dependent probe field is negligible, so we assume  $\dot{p}_{+}(t)=0$. Moreover, we assume that the system is working in the resolved sideband limit ($\kappa < \omega_m$), the Stokes field generated by the interaction of the coupling field with the mirror is very small, therefore we assume $\dot{c}^{*}_{-}(t)=0$. Substituting Eq. \ref{anzat1} into Eq. \ref{new}, we obtain
\begin{eqnarray}
\frac{dq_+}{dt}  &=& \frac{1}{s} \{ d-i\hbar g^2 |c_0|^2 \}q_+
-[\frac{\hbar g}{m(\gamma_m - i\delta)}c_0^\star] c_+ \nonumber \\
\frac{dc_+}{dt}  &=& -[2 \kappa+i(\Delta - \delta)]c_+ - i g c_0
q_+ + \varepsilon_p(t), \label{polariton}
\end{eqnarray}
where $s=m(\gamma_m - i\delta)[2 \kappa - i(\Delta + \delta)]$,
$d=m[2 \kappa-i(\Delta+\delta)](i\delta
\gamma_m-\delta^2-\omega_m^2)$, and $\Delta=\omega_0 - \omega_c +
g \,q_0 $ is the effective detuning. The zeroth order solutions are
$c_0 = \varepsilon_c/(2\kappa+i \Delta)$ and $ q_0 = - \hbar g
|c_0|^2/m \omega_m^2$. Eq. \ref{polariton} describes the coupled, normal mode
excitations of mirror and optical modes
propagation of probe field in a nanomechanical system. One can
solve Eq. \ref{polariton} by introducing the matrix notation
\begin{eqnarray}
\overrightarrow{V}&=& \left( \begin{array}{cc} q_+ \\c_+
\end{array} \right),\label{mtV} \\
 \widetilde{M}&=& \left( \begin{array}{cc} A & B \\C & D
\end{array} \right),\label{mtM}\\
\overrightarrow{F}&=& \left( \begin{array}{cc} 0
\\\varepsilon_p(t)
\end{array} \right),\label{mtF}
\end{eqnarray}
where $A=(-d + i\hbar g^2 |c_0|^2)/s$, $B=\hbar
gc_0^\star/m(\gamma_m - i\delta)$, $C=i g c_0$, and $D=2
\kappa+i(\Delta - \delta)$.  Eq. \ref{polariton} becomes
\begin{equation}
\frac{d}{dt}\overrightarrow{V}=-\widetilde{M}\cdot
\overrightarrow{V} + \overrightarrow{F}(t), \label{polariton1}
\end{equation}
whose solution can be expressed as \cite{scully}:
\begin{equation}
\overrightarrow{V}(t)=e^{-\widetilde{M} (t-t_0)}
\overrightarrow{V}(t_0) + \int_{t_0}^{t} e^{-\widetilde{M}
(t-t^{'})} \overrightarrow{F}(t^{'}) dt^{'}.
\end{equation}
If we take $t_0\rightarrow-\infty$ the solution becomes :
\begin{equation}
\overrightarrow{V}(t)= \int_{-\infty}^{t} e^{-\widetilde{M}
(t-t^{'})} \overrightarrow{F}(t^{'}) dt^{'}. \label{sol1}
\end{equation}
If $\overrightarrow{F}(t^{'})$ is constant the steady state
solution $\overrightarrow{V}=\widetilde{M}^{-1}\overrightarrow{F}$
and $\dot{\overrightarrow{V}}=0$.
\begin{figure}[!t]
\begin{center}
\centering{\vspace{0.5cm}}
\includegraphics[width=0.4\textwidth]{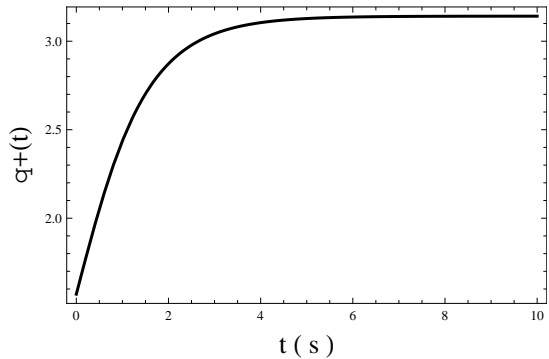}
\caption{\label{fig9} $q_+(t)$ as a function of time with $\Delta=\omega_m$.
All parameters are the same with those of Fig. \ref{fig2}.}.
\end{center}
\end{figure}
%
\begin{figure}[!t]
\begin{center}
\includegraphics[width=0.4\textwidth]{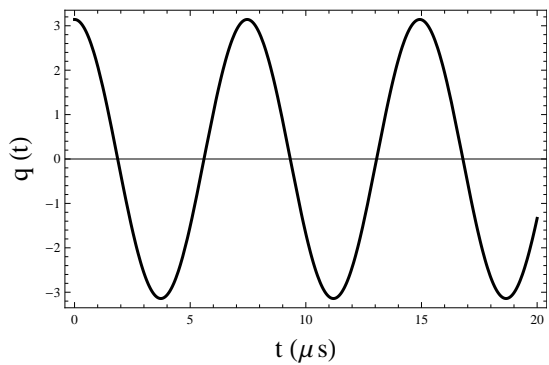}
\caption{\label{fig10} Time dependence of
$q(t)$ for $\Delta=\omega_m$ and  $\delta=\omega_m$. All parameters
are the same with those of Fig. \ref{fig2}.}.
\end{center}
\end{figure}
%
We take the pump field constant, whereas the probe
field depends on time. After solving Eq. \ref{polariton}
analytically, we find $c_+(t)$ and $q_+(t)$ in terms of hypergeometric
functions and the final result is plotted in Fig. \ref{fig9} under the EIT condition of $\Delta=\omega_m$. We
plot mirror vibrations($q_+(t)$) as a function of
time in Fig. \ref{fig9} for  $\varepsilon_p(t)=\textrm{sech}(t)$.
The total displacement of the robust excitations of NMM is:
\begin{equation}
q(t)=q_0+ 2 q_+(t)\cos{\delta t}.
\label{displacement}
\end{equation}

We find situations with both robust excitations of NMM motional modes where the mirror is simply displaced by the optical intensity. This behavior is displayed in Fig. \ref{fig10}. During the final stages of this
work, similar time dependent control
procedures are considered for optomechanical quantum
memory applications
\cite{agarwal2012}.

\section{conclusion} \label{sec:conclusion}
We have examined the question of delay and advance of the probe
field under the conditions of electromagnetically induced
transparency in optomechanical system contained in a high quality
double-ended cavity. We have shown that it is possible to control
the propagation of probe pulse in a double-ended cavity with a
NNM. We have computed the transmission and reflection spectrum of
the probe field. Tunable group delay and advance of optical pulse
by adjusting the pump power are found. As the pump power increases
the group delay becomes smaller, while it saturates beyond a
critical value of the pump power. The magnitude of the group delay
is found to be  $\sim 1\,$ms and the advance is $\sim -2\,$s at a
low pump power $\sim 0.2\, \mu$W for the parameters chosen as in
Ref. \cite{thompson}. The system under consideration is easier to
implement and offers longer group delays in comparison to other
optomechanical proposals \cite{kadizhu}. Moreover, we have
investigated the interaction of the optomechanical cavity with a
time dependent probe field for controlled excitations of mirror
vibrations and therefore, we have showed that thanks to optical
intensity, mechanical mode excitations can be  caused by a time
dependent probe field.
\begin{acknowledgements}
D.T. acknowledges support from Faculty Support Program by the
Council of Higher Education (Y\"{O}K) of Turkey. D.T thanks to
G. S. Agarwal for many stimulating and fruitful
discussion and his
hospitality at the Oklahama
State University, Stillwater, USA.
\end{acknowledgements}

\end{document}